\documentclass{IEEEcsmag}

\usepackage[colorlinks,urlcolor=blue,linkcolor=blue,citecolor=blue]{hyperref}

\usepackage{upmath}
\usepackage{caption}
\usepackage{subcaption}
\usepackage{tikz}
\usepackage{graphicx}
\usepackage{amsmath}
\usepackage{gensymb}
\usepackage{amsfonts}
\usepackage{amssymb}
\usepackage{mathtools}

\usepackage[utf8]{inputenc}
\usepackage{tabularx}
\usepackage{booktabs}
\usepackage{amsmath}
\usepackage[version=3]{mhchem} 
\usepackage{siunitx}
\jvol{XX}
\jnum{XX}
\paper{8}
\jmonth{May/June}
\jname{arXiv}
\pubyear{2022}

\begin{document}


\title{DNA Storage: A Promising Large Scale Archival Storage?}

\author{Yixun Wei}
\affil{Department of Computer Science and Engineering, University of Minnesota, Minneapolis, Minnesota, USA}

\author{Bingzhe Li}
\affil{Department of Electrical and Computer Engineering, Oklahoma State University, USA}

\author{David H.C. Du}
\affil{Department of Computer Science and Engineering, University of Minnesota, Minneapolis, Minnesota, USA}


\begin{abstract}
Deoxyribonucleic Acid (DNA), with its high density and long durability, is a promising storage medium for long-term archival storage and has attracted much attention. Several studies have verified the feasibility of using DNA for archival storage with a small amount of data. However, the achievable storage capacity of DNA as archival storage has not been comprehensively investigated yet. Theoretically, the DNA storage density is about 1 exabyte/mm$^3$ (10$^9$ GB/mm$^3$). However, according to our investigation, DNA storage tube capacity based on the current synthesizing and sequencing technologies is only hundreds of Gigabytes due to the limitation of multiple bio and technology constraints. This paper identifies and investigates the critical factors affecting the single DNA tube capacity for archival storage. Finally, we suggest several promising directions to overcome the limitations and enhance DNA storage capacity.

\end{abstract}

\maketitle

\chapterinitial{INTRODUCTION} An indisputable fact is that we are surrounded by more and more available data. Meanwhile, various applications such as IoT, autonomous vehicles, video streaming, scientific discovery, etc. need to preserve the increasingly large available datasets for future data analysis/mining in a longer duration. 
Unfortunately, the development of storage technologies/devices is not matching with the booming storage demand. Moreover, traditional storage media are not durable enough (e.g., disks for 3-5 years, tapes for 10-15 years) for the increasingly longer data retention duration.  Therefore, data must be migrated from the existing obsolete devices to new devices when the current devices wear out every few years. The data migration and device replacement cost can be over half of the total cost of ownership even in cloud archival storage~\cite{AA1}. To meet the burgeoning storage demand, we are looking for a new storage medium with high density and long durability.

DNA has emerged with its ultra-high storage density (i.e., a theoretical storage density of about 1 exabyte/mm$^3$~\cite{ASPLOS}). When keeping DNA in a proper environment, it can last for centuries~\cite{AA1} and thus save an enormous amount of maintenance cost for data transfer and device replacement. As DNA synthesis/sequencing (i.e., write/read) prices keep decreasing, DNA storage is identified as a promising archival storage medium.

Several researchers have demonstrated the feasibility of DNA archival storage. Their investigations focused on error-tolerant encoding schemes~\cite{FEC}\cite{Fountain}\cite{GF47}\cite{IMGDNA} and random accessibility with wet-lab experiments of a small amount data~\cite{ASPLOS}\cite{Random}. Encoding schemes are responsible for the conversion between digital data and DNA sequences. Random access allows retrieval of target data without sequencing the entire DNA pool which greatly saves the number of slow and expensive sequencings needed. However, the achievable storage capacity of using DNA for archive based on current biotechnologies remains unknown. In DNA storage, Data is in the form of powder for a longer duration. When sequencing, people liquidize the powders and use a drop of the liquid for sequencing. To avoid long diffusion time and ensure sequencing effectiveness, people usually separate a large powder mixture into multiple physical tubes. A tube is a basic DNA storage device, and DNA storage capacity can be measured by DNA tube capacity. A larger single tube capacity means larger overall DNA storage capacity or higher storage density. In this paper, we study the storage capacity of a single tube.


One important aspect of limiting the real DNA storage capacity is that DNA storage is error-prone. Due to the mutation nature of DNA, errors like insertion, deletion, and substitution can happen during synthesis and sequencing processes. Some special DNA patterns may increase the probabilities of these errors such as long homopolymer, secondary structure, etc. Therefore, it is important to avoid certain biological constraints to decrease the error rate of DNA storage using special DNA encoding schemes. We also convert digital data to error-tolerant DNA sequences with a shorter DNA strand length (the current practical strand length for DNA storage is between 200 to 300 nt) since the longer the DNA strand length, the higher the error rate incurred. 



For DNA storage used for archive, we consider random access is a must. The most prevalent sequencing technology, Polymerase Chain Reaction (PCR), may also introduce extra errors. PCR-based random access is a process that amplifies only target DNA strands to enough concentration to be sequenced out. It uses a pair of short synthetic nucleotide sequences called primers to tag a set of DNA strands (the number of DNA strands in a set is considered as a parallel factor since these DNA strands can be read out in one sequencing process). Primers should be sufficiently different from each other such that random access can accurately retrieve the target strands. However, the requirement of sufficient difference between primers reduces the number of usable primers. Similarly, the encoded DNA strands must not contain certain DNA short sequences that are also included in primers. This will further limit the DNA storage tube capacity. We will discuss these capacity limitations in the next section.

In this paper, we investigate the single tube capacity of DNA archival storage. We first list possible factors that influence DNA tube capacity. We then generate a large enough primer library to measure the practical tube capacity with different encoding schemes and different types of data. Due to the extremely high synthesis/sequencing cost for a large amount of data at the current time, the measurements are done without wet labs. We follow a similar setup in existing studies including the value of the parallel factor, bio-constraints, collision definitions, etc., and use NCBI BLAST~\cite{BLAST}, one of the most widely used genomic sequence analysis tools, to check the feasibility of DNA storage. According to our measurements, DNA tube capacity can only reach hundreds of gigabytes. Finally, based on the thorough analysis of the factors restricting tube capacity, we suggest several potential ways to improve the DNA tube capacity.
 


The rest of this paper is organized as follows. We first discuss the necessary background. Then we discuss several capacity limitations of DNA archival storage. After that, we present our measurements of the practical DNA tube capacity. Finally, we propose several potential enhancement approaches to enlarge DNA tube capacity.

\section{BACKGROUND}

\begin{figure*}[htbp]
\centering
\includegraphics[width=\textwidth,height=4.5cm]{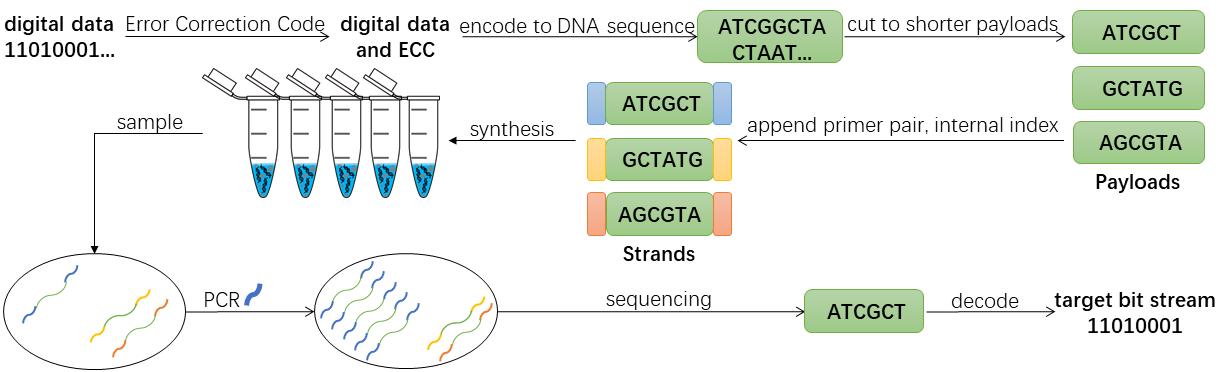}
\caption{Workflow of DNA Synthesis and Sequencing}
\label{fig:DNA storage}
\end{figure*}

\noindent This section explains basic background of DNA storage synthesis and sequencing workflows and PCR procedure for random accesses. 

\subsection{DNA Storage Synthesis/Sequencing Workflows}

\noindent{\bf Figure 1} summarizes the workflows of synthesis and sequencing of DNA storage. A DNA storage system takes digital data as input. It first adds Error Correction Codes (ECC)~\cite{Luis} to digital data to provide recoverability in case of existing synthesis/sequencing errors. It then coverts the digital data into a sequence of nucleotides based on an encoding scheme and cuts the sequence into multiple shorter sequences called payloads. After that, it flanks a primer pair and an internal index onto each payload sequence to form a DNA strand. The primer pair indicates and enables the beginning and end of amplification during the PCR process. Since one primer pair can be shared among multiple DNA strands, an internal index is used to identify the order of DNA strands that share one primer pair. Each DNA strand will be chemically synthesized nucleotide by nucleotide and stored in physical tubes/pools. To recover digital data back, target DNA strands with a specific primer pair can be selectively amplified via PCR and sequenced via next-generation sequencing (NGS) or other sequencing techniques. Finally decoded back to digital information.

\subsection{PCR Procedure}
\noindent PCR is a method that can amplify target DNA strands to enough concentration for DNA sequencing. It enables random access of target data without sequencing the whole storage pool. A PCR process usually consists of 25 to 35 cycles. Each cycle doubles the number of target strands with the specific primer pair. As shown in {\bf Figure 2(a)}, a standard PCR cycle follows three steps: denaturation, annealing, and elongation. In the denaturation step, DNA strands are denatured into two single-stranded DNA molecules. In the annealing step, complementary primers will be added to the reaction solution and bind to the target denatured DNA molecules. In the elongation step, with the help of dNTPs, the Taq DNA polymerase adds nucleotides to the growing DNA strand.
\begin{figure}[htbp]
  \begin{subfigure}[b]{0.45\textwidth}
    \includegraphics[width=\textwidth]{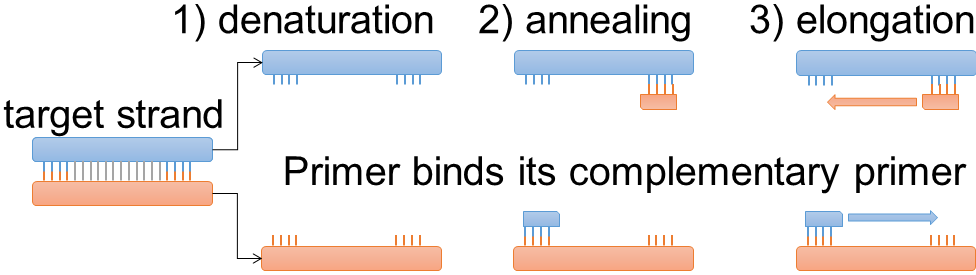}
    \caption{standard PCR}
    \label{fig:standard PCR}
  \end{subfigure}
  \hfill
  \begin{subfigure}[b]{0.45\textwidth}
    \includegraphics[width=\textwidth]{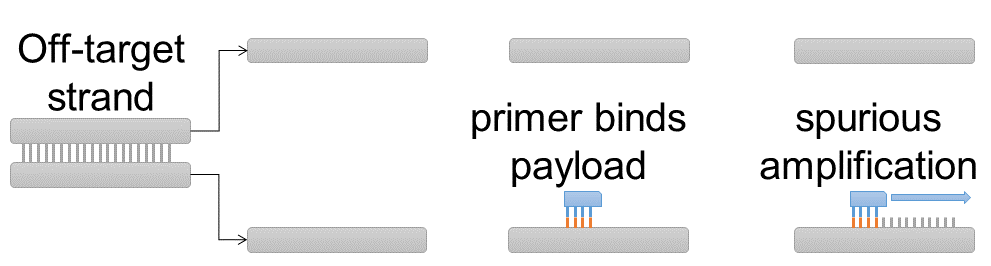}
    \caption{PCR with primer-payload collision} 
    \label{fig:primer-payload collision}
  \end{subfigure}
  \caption{standard PCR and defective PCR with primer-payload collision}
\end{figure}

\section{Factors Affecting DNA storage capacity}
\noindent Typically, the single tube capacity of PCR-based DNA storage is affected by the parallel factor, payload encoding density, payload length, and the number of usable primers.

\textbf{Parallel factor} is the number of strands that share a primer pair. A larger parallel factor produces a larger capacity but requires more PCR cycles to amplify all strands to enough concentration. An overlarge parallel factor can cause data loss if some strands fail to be amplified to enough concentration. Existing studies have adopted  $1.55 \times 10^6$ as a practical parallel factor~\cite{FEC}\cite{Li2020can}. Parallel factor together with the number of useable primers can determine how many DNA strands per tube. 

\textbf{Payload encoding density} is the number of bits each nucleotide can represent. To reduce the potential errors during synthesis/sequencing, an encoding scheme needs to avoid specific patterns such as long homopolymer, low/high GC content ratio, etc. As a result, an ideal encoding density, 2 bits/nt, cannot be achieved/used.  Encoding schemes are required to convert digital data without generating long homopolymers and unbalanced GC content. A theoretical upper bound of encoding density is proofed around 1.98~\cite{Fountain}. 
    
\textbf{Payload length} is the length of a DNA strand subtracting the length of primer pair, internal index, and logical redundancy (e.g., ECC). The overall DNA strand length is currently restricted by synthesis technology. A practical DNA strand length is usually no more than 300 nt. Otherwise, it is likely to have more synthesis errors~\cite{Luis}. 15\% redundancy is usually enough to recover data from errors~\cite{Luis}. Therefore, excluding redundancy, internal index and primer pair, payload length in a DNA strand can be varied around between 100 nt - 250 nt.

\textbf{The number of usable primers} refers to how many primers can be safely used for PCR-based random access in a single tube. Theoretically, with the primer length of 20 bases, the number of primers can reach up to $4^{20}$. However, due to bio-constraints, the number of primers used in a DNA storage tube is much less than expected. The bio-constraints include homopolymer length shorter than 4, balanced GC content, and the absence of secondary structures. Moreover, since primers serve as unique tags in PCR, they must be sufficiently different from each other to avoid potential crosstalk. The details are shown in the next section when we discuss building a primer library with enough primers.

Even if primers are designed orthogonal to each other, they can still interact with payloads, which are referred to as primer-payload collisions~\cite{Random}. A primer-payload collision happens when a primer and a payload have a pair of almost identical subsequences. The pair of almost identical subsequences are usually longer than 12 bases and have at most two mismatches or gaps. In each PCR cycle, wrong annealing can happen if any payload has a collision with the target primer. The wrong annealing will amplify some irrelevant payloads and, more importantly, consume the limited PCR reagents (e.g., complementary primers used to bind with the target primer and free DNA bases waiting for complementing new double-helix strands). 

\begin{figure}[!t]
\centering
\includegraphics[width=0.48\textwidth]{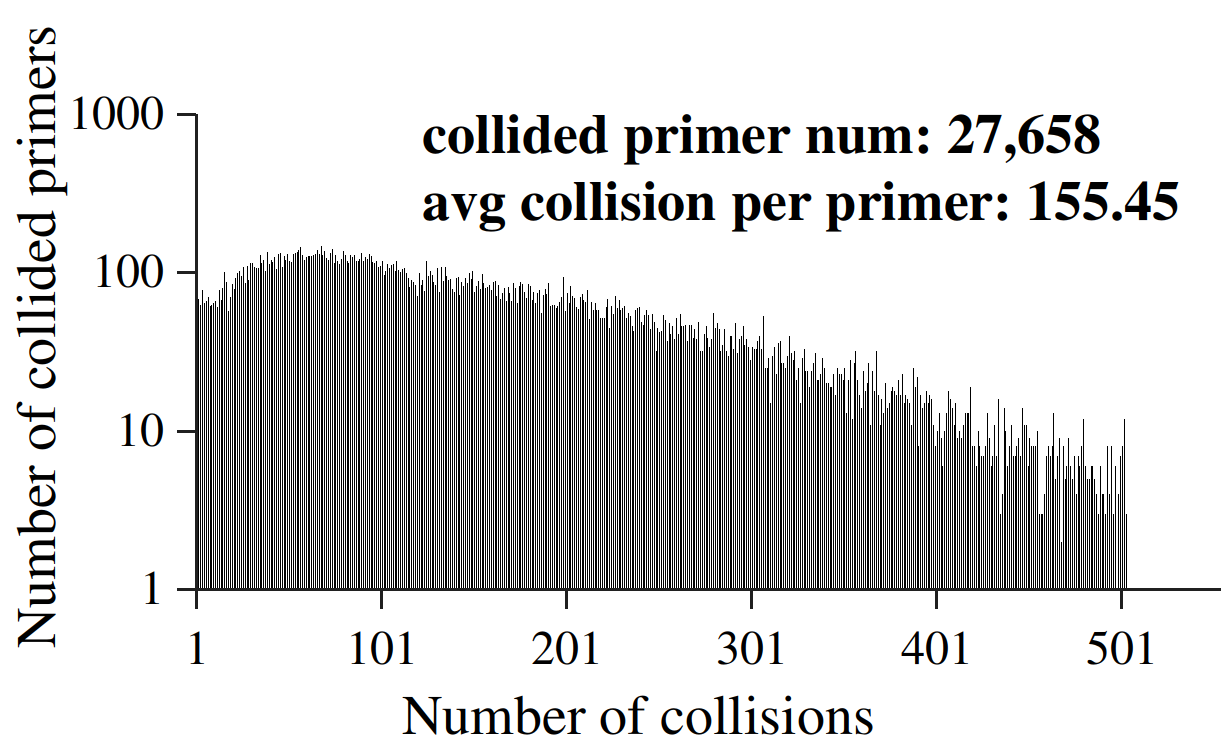}
\caption{Distribution of primers with different numbers of collisions (encoding scheme: Blawat, data: 135MB video)}
\label{fig:collision distribution}
\end{figure}

Figure~\ref{fig:collision distribution} shows the distribution of primers with different numbers of collisions. The input data is a 135MB video (\text{\url{https://www.youtube.com/watch?v=qybUFnY7Y8w}}), and the encoding scheme is Blawat code~\cite{FEC} (see the encoding scheme, primer generating, and collision checking details in next section). Although input data is only 135MB, 27,658 primers (i.e., 98.77\% primers in our primer library) have collisions with those payloads, and the average number of collisions per collided primer is 155.45. If the input data keeps scaling up, most collided primers likely have hundreds or thousands of collisions. In this case, PCR amplification is very vulnerable to primer-payload collisions. The target primer must compete for the limited PCR reagents with thousands of collided payloads in each PCR cycle. Even if a primer has only a few collisions at the beginning, the collisions can still significantly impact the final result. That is because PCR is an exponential augmentation process. Slight amplification variations can exacerbate cycle by cycle and consume ever more PCR reagents. Eventually, the resultant solution will be saturated with plenty of irrelevant sequences but very few target sequences. As a result, the sequencing for target sequences is inhibited.

Currently, the collision problem has not drawn enough attention for two reasons. Firstly, existing works are mostly based on a small scale data (i.e., several Kilobytes to several Megabytes)~\cite{Church}\cite{GF47}\cite{ASPLOS}\cite{FEC}\cite{Fountain}. Secondly, the data in DNA pool is not diverse enough. Existing works build a DNA pool by mixing a few synthetic target data with background non-target data. The non-target data are usually homogenous DNA strands made from PCR amplification because PCR amplification is much cheaper than DNA synthesis. Because non-target data is almost identical, it has less chance of colliding with primers during PCR. However, DNA payloads in a tube should be more diversified when DNA storage is put to practical use and continues to scale up in capacity. In this case, the impact of primer-payload collisions will be very severe. For example, a previous study~\cite{Random} reported a PCR failure because of primer-payload collisions. They expanded a DNA tube from a single file to multiple files and failed to sequence target data when there were nine files in the tube (the target data is 17.4\% of the whole data in the tube). To ensure a precise and effective PCR, it is essential only to use primers that have no collisions with payloads.

Given parallel factor, payload length, and payload encoding density, we can calculate the amount of data each primer pair can accommodate. Since the above three factors are all restricted by the current biological technologies, the number of usable primers becomes a major factor in deciding practical DNA tube capacity. Since the primer-payload collision is input data dependent, we plan to use real-world data as input to check how many primers are usable and what is the practical tube capacity. 

\section{Investigation of Practical Tube Capacity}
\label{sec:4}
\noindent This section illustrates the impact of primer-payload collisions and discovers the practical tube capacity.

\subsection{Build A Large Primer Library}
\label{sec:primer rules}
\noindent We first build a large enough primer library since the existing primary libraries are too small for our purpose. We adopt a primer design framework designated for DNA storage to generate our primer library~\cite{Random}. The framework keeps generating random 20 base DNA sequences and evaluating the sequences with a set of criteria. Sequences can be added to the primer library only if they satisfy all the following criteria:

\begin{itemize}
    \item GC content $\in$ [0.45, 0.55]
    \item melting temperature $\in$ [55$^\circ$C, 60$^\circ$C]  
    \item homopolymer length $\leq$ 3
    \item Hamming distance between any two primers $\geq$ 6
    \item no more than 10 bases inter-complementarity 
    \item no more than 4 bases intra-complementarity
\end{itemize}

Because GC contributes more bonding power than AT, proper GC content is required to ensure stable binding of primers. Similarly, melting temperature, the temperature at which DNA duplex dissociates to be single-stranded, is also required for reliable primers binding. Hamming distance is required to ensure primers are sufficiently different. It helps reduce the likelihood of non-specific primer binding. 

Intra-complementarity refers to two subsequences within a primer complementing each other in reverse order. Similarly, inter-complementarity refers to two complementary subsequences in two primers. More than 4 bases intra-complementarity or 10 bases inter-complementarity can cause a primer hybridizes with itself (i.e., hairpin) or with other primers (i.e., primer dimer) instead of annealing to the binding site of the target payloads. For example, primer "...ATGG...CCAT..." has intra-complementary subsequences with length 4. 

This primer design framework has been verified by Organick et al.~\cite{Random} with a pool of 3,240 small synthetic files. They successfully retrieved up to 48 files in a one-pot multiplex PCR experiment. They also predicted that the number of potentially generated primers is roughly 28,000. We follow their design framework in this study to generate a primer library of 28,000 primers. More primers may be available with different primer design rules, but 28,000 primers are enough for us to investigate the capacity loss due to primer-payload collisions. Hereafter, the practical tube capacity refers to the capacity with the primer library of 28,000 primers and is based on the current biotechnologies.

\subsection{Encoded Digital Data}
\label{sec:encoding}
\noindent We collect 1,500 GB of data with five data types, including image, video, audio, eBook, and software archives (300 GB of each type). Image data is a subset of \texttt{ImageNet}~\cite{Imagenet}. Audio data contains 187GB of \texttt{LibriSpeech}~\cite{LibriSpeech} and some small collections from \texttt{InternetArchive} (\text{\url{https://www.youtube.com/watch?v=qybUFnY7Y8w}}). For other data types, we do not find any single dataset that is large enough and thus collect multiple small datasets from \texttt{InternetArchive}. 

We implement the following five highly cited DNA encoding schemes. For some encoding schemes that have no legit name, we name them using the first author's name.  

\begin{itemize}

\item\textbf{Church}~\cite{Church} is one of the earliest encoding scheme. It encodes one bit to one base, and every bit has two optional bases to choose from (i.e., 0 to AC and 1 to TG). It selects proper bases to represent digital bits, so there are no overlong homopolymers or unbalanced GC content. Its encoding density is 1 bit/base.
    
\item\textbf{Rotation} code has been used in multiple DNA storage studies~\cite{ASPLOS}\cite{Random}\cite{IMGDNA}\cite{Goldman}. It first converts binary data to ternary. Then it encodes the current ternary digit to a specific base depending on the last encoded base. The current ternary digit can only be encoded to a base different from the last one. Rotation code has no homopolymers, and its encoding density is 1.58 bits/base.
    
\item\textbf{Blawat}~\cite{FEC} encodes every eight binary bits to five bases. Bits 0 \& 1, 2 \& 3, and 4 \& 5 are respectively encoded to the first, second, and fourth bases, while bits 6 \& 7 decide the third and fifth bases. The encoding of bits 0-5 is a direct conversion from \{00,01,10,11\} to \{A,C,G,T\}. While the encoding from bits 6 \& 7 to the third and fifth bases are selected from a mapping table: $4^2$ possible two-bases combinations map to $2^2$ possible bit patterns. The third base will be selected to ensure the first three bases will not be identical, and the fifth base will be selected to ensure the last two bases will not be identical. With these constraints, the longest possible homopolymer length is 3, and the encoding density is 1.6 bits/base.

\item\textbf{Grass}~\cite{GF47} first converts binary bits to Galois Field of size 47 (GF(47)). It then bases on a mapping table to encode GF(47) digits to base triplets in which the second base differs from the third base. Since no base triplet has identical second and third bases, the longest homopolymer length in the resultant DNA sequences is three. The encoding density of Grass encoding is 1.77 bits/base.

\item\textbf{Fountain} code is famous for its robustness~\cite{Fountain}. It iterates over Luby transform to select data segments and XOR them bit-wisely. The resultant data will be directly translated to DNA sequences from \{00,01,10,11\} to \{A,C,G,T\}. Then, it screens out the sequences with undesired GC content and long homopolymers. It keeps the Luby transform until generates a predefined number of DNA sequences. For reliability, the number of resultant DNA sequences is recommended to be 5\%-10\% more than the total input segments (in our case: 10\%). Its encoding density is 1.81 bits/base if we count the overhead of the 10\% redundancy.
\end{itemize}

In the encoding process, we first cut the collected five types of digital data into multiple small files of 10 MB. We will add files incrementally for collision checking and capacity identification later. We further partition each file into chunks of 239 bytes and apply Reed-Solomon code RS(255,239) to append 16 parity bytes to each chunk (i.e., each chunk has a size of 255). After that, we use each of the five encoding schemes to encode data chunks to DNA sequences and further partition these sequences into multiple DNA payloads with a length of 198\footnote{For the sake of alignment, the last chunk of each 10MB file and the last payload of each chunk will have different size.}. We chose 198 because it is in the middle of the payload length range and is dividable by three, which is required by Grass code. The DNA payloads will further be appended with an internal index of 12 bases to indicate their ordering associated with a primer pair. We choose 12 bases for internal index length because it is long enough to generate more than $1.55 \times 10^6$ base sequences that will not collide with any primers in the primer library. We will choose $1.55 \times 10^6$ of them in a lexicographical order to form a mapping table to represent the ordering of DNA payloads associated with a primer pair. The payloads with internal indexes will be sent to BLAST to check possible collisions. When decoding DNA sequences back to digital data, the mapping table can be used to identify the order of payloads. 

\subsection{Check Collisions and Discover Tube Capacity}

After building the primer library and encoding digital data to DNA payloads, the next step is filtering out primers with primer-payload collisions. Unfortunately, synthesizing such a large scale of data for collision checking is currently cost prohibitive (predicted at \$21,000/MB by 2027~\cite{DNAfuture}). Instead, we follow Organick's work \cite{Random} and use BLAST~\cite{BLAST} to check collisions between primers and payloads. Only primers without any collisions will be considered as usable primers. 

We check collisions between all kinds of DNA payloads gathered in the last subsection and all primers from our primer library. Payloads are in multiple partitions and each partition represents 10.66MB input data (i.e., 10MB digital data + 0.66MB ECC). We sequentially check collision on each payload partition (i.e., increase of input data) and accumulate the collided primers. Meanwhile, more usable primers are needed to accommodate data as input data increases. The number of usable primers required to store a data set is the total data size divided by the encoding density, payload length, parallel factor, and two (i.e., two primers to form a primer pair). We keep increasing the input data until the number of remaining usable primers decreases to a number that is just capable of accommodating all input data. We consider the size of digital data in the input data (i.e., 10MB $\times$ number of input files) as the practical tube capacity of a specific data type for a particular encoding scheme. 

\begin{figure}[!t]
\centering
\includegraphics[width=0.48\textwidth]{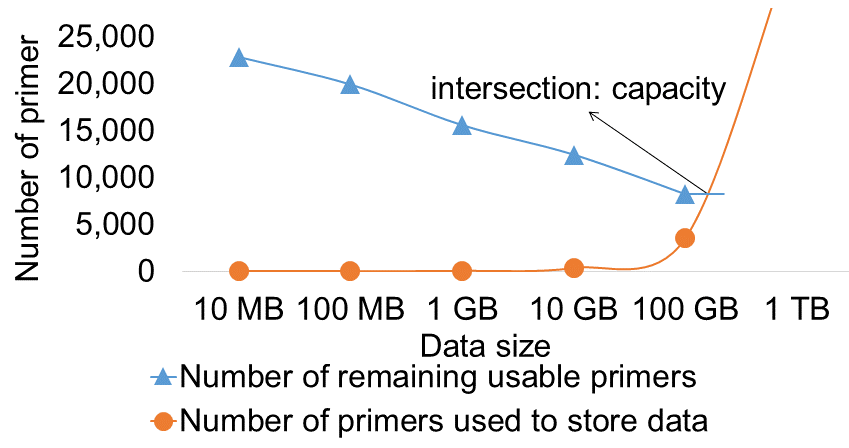}
\caption{Practical tube capacity as increasing input data size using Rotation code. The input data is ImageNet.}
\label{fig:Practical tube capacity}
\end{figure}

Figure~\ref{fig:Practical tube capacity} shows the achievable tube capacity when use Rotation code and \texttt{ImageNet}. The orange line in Figure~\ref{fig:Practical tube capacity} shows the increased number of primers needed to accommodate the input data. The blue line shows the decreased number of remaining usable primers. When input digital data grows to 215.42GB, the remaining usable primers are just enough to accommodate the input data. Due to space limitations, we do not show the figures of all five encoding schemes with five data types. The number of usable primers for other encoding schemes drops quickly, even with small amount of input data. We summarize all the results in the next subsection.

\subsection{Summary of Tube Capacity}

\begin{table*}[htb]
\centering
  \caption{Practical DNA tube capacities for different encoding schemes with multiple data types considering primer-payload collision.\label{tab:capacity}}
  \begin{tabular}{l*{8}{c}}
    \toprule
    & \multicolumn{1}{c}{\textbf{Encoding Density}} & \multicolumn{1}{c}{\textbf{Capacity without}} &  \multicolumn{5}{c}{\textbf{Practical Achievable Capacity (GB)}} \\ \cmidrule(){4-8}
    & {\textbf{(bits/base)}}  & {\textbf{considering collisions (GB)}}
    & {\textbf{Image}} & {\textbf{Audio}} & {\textbf{Video}} & {\textbf{eBook}} & {\textbf{Software}} \\
    \midrule
        Church & 1  &  468.80  & 0.14  &    0.19  &    0.18  &    0.15  &    0.18  \\
        Rotation & 1.58  &  740.71 & 215.42  &   220.82  &    218.71  &    211.41  &    217.20 \\
        Blawat & 1.6  &  750.08  & 1.51  &  1.93  &    1.71 &    1.65  &    1.35 \\
        Grass &  1.78 &   834.46 & 6.26  &  6.87 &    6.72 &    6.00  &    6.27 \\
        Fountain &  1.81  &  848.53  & 0.41  &   0.95  &    0.59 &    0.37  &    0.44  \\
    
    \bottomrule
  \end{tabular}
\end{table*}

Table~\ref{tab:capacity} shows the practical, achievable tube capacity of the five encoding schemes with five types of data. Suppose all primers in the library are usable without primer-payload collisions. In that case, the five encoding schemes will achieve a tube capacity varying from 468GB to 848GB, depending on their encoding densities. However, if we remove the primers that have collisions, Blawat/Fountain/Church/Grass codes will suffer more than 99\% practical tube capacity reduction and can have only hundreds of Megabytes to a few Gigabytes capacity. Rotation code shows the most collision resistance but still suffers 70\% capacity reduction and achieves a capacity of around 215GB.  

Blawat/Fountain/Church codes are close to direct conversion between binary data and DNA sequences (from {00, 01, 10, 11} to {A, C, G, T}). Since no special pattern is applied, the resultant DNA sequences are possible to collide with most primers. An example of the distribution of primer-payload collisions is shown in Figure~\ref{fig:collision distribution}. The encoding scheme is Blawat code, and the input data is the 135MB video. Despite the small data set, Blawat code has already collided with 27,658 primers, and the average number of collisions per collided primer is 155.45. However, Blawat code is still a little better than Fountain and Church codes because Blawat code only directly converts the first 6 bits of every 8 bits. Besides, Fountain code performs slightly better than Church code because of its higher encoding density. 

By contrast, Rotation code has the second lowest encoding density but achieves the highest tube capacity among the five encoding schemes. That is because Rotation code sacrifices its encoding density to avoid consecutive identical bases (i.e., no homopolymers). Its DNA payloads are less likely to collide with primers that have consecutive identical bases. As for Grass code, it encodes digital data with base triplets in which the second base is different from the third base. This makes Grass's DNA payloads have less number of three consecutive identical bases and thus less likely to collide with primers that have three consecutive identical bases. However, Grass's DNA payloads still collide with primers that have shorter consecutive identical bases. Therefore, Grass code is worse than Rotation code but better than Blawat/Fountain/Church codes. Besides, the collision problem exists in all five data types, but no significant capacity differences were observed among different data types. That may be because different data types will eventually cover most of the possible digital sequences as the data size increases.





\section{Potential Capacity Enhancement}
\noindent Although several factors like strand length and parallel factor are temporarily limited by the current technologies, several potential approaches can enhance DNA tube capacity for archive.

First, we can investigate new primer design frameworks to generate more primers. The existing primer frameworks are designed for biological constraints rather than the need for more usable primers. They apply strict rules on primer design to ensure PCR specificity. A primer design framework can more aggressively limit some primer design space to ensure both primer reliability and produce more usable primers. 

Besides generating more usable primers, we can also work on the payload side. Payload sequences in DNA storage are synthetically generated based on an encoding scheme. Given a primer library, it is possible to design special encoding schemes to encode DNA sequences by avoiding primer-payload collisions. The previous section has shown that the Rotation scheme can avoid some collisions via tightening the longest homopolymer length from 3 to 1. Besides homopolymer length, we can further create other DNA sequence patterns to be different from those of primers. For example, we can tighten DNA sequence to have GC content $\in$ [0.45, 0.50] to avoid collisions with primers that have GC content $\in$ (0.50, 0.55]. That is, we can co-design primers and encoding schemes together. 
However, special encoding schemes that create different patterns may lose some encoding density. That is because a smaller number of DNA sequences are qualified to represent digital data. 

Another cost-efficient way is to use some payload post-processing.  If we can have variable-length payloads and make them break up at the collision locations, we can remove many collisions. If we can change payload content with a simple permutation of nucleotides, we can also remove a considerable number of collisions. These potential approaches work in different aspects and can be applied together to enhance DNA storage capacity. 

\section{Conclusion}  

In this paper, we identify the limitations of DNA tube capacity based on the-state-of-the-art synthesis and sequencing technologies. We then build our own large enough primer library and investigate the tube capacity of different encoding schemes with different types of data. The results show that the tube capacities of the current methods are reduced to hundreds of gigabytes, which is far less than what people expected. By analyzing the DNA storage capacity limitation and bio-constraints, we point out some potential ways to enhance DNA storage tube capacity. We hope with further investigations of these potential approaches and the development of new bio-technologies, DNA tube capacity will be sufficiently large for practical archival storage.

\end{document}